\documentclass[twocolumn,showpacs,preprintnumbers,amsmath,amssymb]{revtex4}

\usepackage{graphicx}
\usepackage{dcolumn}
\usepackage{bm}
\usepackage{epsfig}

\begin{document}

\preprint{Submitted to Phys. Rev. B }

\title{Pinning and creation of vortices in superconducting films by a magnetic dipole }
\author{Gilson Carneiro}
\affiliation{Instituto de F\'{\i}sica, Universidade Federal do Rio de Janeiro,  
C.P. 68528, 21941-972, Rio de Janeiro-RJ, Brasil }
 \email{gmc@if.ufrj.br}

\date{\today}

\begin{abstract}

Interactions between vortices in planar superconducting films and a point magnetic dipole placed outside the film, and the creation of vortices by the dipole, are studied in the London limit. The exact solution of London equations for films of arbritrary thickness with a generic distribution of  vortex lines, curved or straight, is obtained by generalizing the results reported by the author and E.H. Brandt ( Phys. Rev. B {\bf 61}, 6370 (2000)) for films without the  dipole.  From this solution the total energy of the vortex-dipole system is obtained as a functional of the vortex distribution. The   vortex configurations created by the dipole  minimize the energy functional. It is shown that the  vortex-dipole interaction energy is given by $- {\bf m}\cdot {\bf b^{vac}}$, where  ${\bf m}$ is the dipole strength and $ {\bf b^{vac}}$ is the magnetic  field of the vortices at the dipole position, and that it can also be written in terms of a magnetic pinning potential acting on the vortices. The properties of this potential are studied in detail.  Vortex configurations created by the dipole on films of thickness comparable to the penetration depth are obtained by discretizing the exact London theory results on a cubic lattice and minimizing the energy functional using a numerical algorithm based on simulated annealing. These configurations are found to  consist, in general, of curved vortex lines and vortex loops. 

\end{abstract}
\maketitle


\section{introduction}
\label{sec.int}

The  experimental  study of artificial superconductor-ferromagnet systems  has received a great deal of attention lately. The magnetic, superconducting, and transport  properties of a great variety of such systems, mostly  superconducting films with  arrays of magnetic dots or anti-dots placed in its vicinity,  have been reported in the literature \cite{expp1,expp2,rev1,expp3}.
The main interest in this type of system is to enhance and modify pinning of vortices, and thereby increase the critical current   and stabilize new vortex phases. 

From the theoretical point of view, these systems are also interesting because they alow  theoretical predictions to be tested in detail experimentally,  since  the ferromagnetic  structures responsible for vortex pinning are well characterized, and can be changed  in a controlable way over a wide range of parametrs.  The goal of theoretical work  is to understand how the presence of the ferromagnet changes the equilibrium and non-equilibrium behavior of vortices. 

The interaction between  vortices and the ferromagnet  results from the action of the inhomogeneous magnetic field created by the ferromagnet on the   superconductor. This interaction is  expected not only to pin vortices placed in the film  by an aplied  field, but also to create vortices, and even to destroy superconductivity in some regions of the sample. The theoretical problem posed by these systems is rather complex, since the equilibrium vortex states in the absence of an applied field are non-trivial. The first problem that  needs to be solved is to calculate the vortex-ferromagnet  interaction and to obtain the equilibrium vortex state resulting from the competition between it and vortex-vortex interactions.  This paper solves this problem for a simple model consisting of a point magnetic dipole placed outside a planar superconducting film of arbitrary thickness in the London limit. First, the exact solutions of London equations for a film with a given distribution of vortices, consisting of a generic  arrangement of straight or curved vortex lines, is obtained.  Second, the total energy of the vortex-magnetic dipole system is calculated as a functional of the vortex distribution. The equilibrium vortex configurations generated by the magnetic dipole can then be  obtained from the energy functional by minimizing it  with respect to the vortex distribution. 

The  main new results reported in  this paper are: i) The proof that the vortex-magnetic dipole interaction energy is  $- {\bf m}\cdot {\bf b^{vac}}$, where ${\bf m}$ is the magnetic moment and ${\bf b^{vac}}$ the magnetic field caused by the vortices at the dipole position,  and that this energy can also be expressed in terms of a magnetic pinning potential for vortex lines of any shape. ii) The exact expression for the energy of the vortex-dipole system  as a functional of the vortex distribution. iii) Approximate vortex configurations generated by the dipole in  films of finite thickness. 

Earlier work on the above described model is restricted to the calculation of the interaction between a magnetic dipole and straight vortex lines. For a semi-infinite superconductor this interaction was obtained by Coffey \cite{coff} as $- {\bf m}\cdot {\bf b^{vac}}/2$. Thin films were considered by Wei et. al. \cite{wei},  by \v{S}a\v{s}ik and Hwa \cite{sah}, and more recently by Erdin et al. \cite{erd1}. The authors of  Refs.\ \onlinecite{wei,sah} find the that the interaction energy  is also $- {\bf m}\cdot {\bf b^{vac}}$, whereas those of  Ref.\ \onlinecite{erd1} obtain the interaction as $- {\bf m}\cdot {\bf b^{vac}}/2$ plus an extra term. In  Refs.\ \onlinecite{wei,erd1,erd2} the creation of vortices by the dipole was investigated by minimizing the energy for simple configurations. Recently, 
Milo\v{s}evi\'c, Yampolsky and Peeters \cite{myp} obtained the energy of interaction between  straight vortex lines and a point magnetic dipole for films af arbitrary thicknesses. These authors find that the  interaction energy consists of two terms: one is  $- {\bf m}\cdot {\bf b^{vac}}/2$  and the other is an interaction between the screening current generated by the dipole and the vortex. These authors also study the creation of vortices by the dipole by examining the interaction energy of several configurations of straight vortex lines.

This paper goes beyond these earlier results by considering the interaction between the magnetic dipole and vortex lines of any shape. This is necessary in films that are not too thin, because the vortex-magnetic dipole interaction is limited to a distance of the order of the penetration depth from the film surface nearer to the dipole, whereas the vortex line energy grows with the film thickness. Thus, creation of straight vortex lines in  thick films  is energetically disfavored.

To solve  London equations for the vortex-dipole system this paper starts from the results obtained by the author and E.H. Brandt \cite{gcehb}  for  films of arbitrary thickness without the dipole.  In  
Ref.\ \onlinecite{gcehb} the magnetic field and energy of an arbitrary vortex distribution in the film are obtained by solving London equations by the method of images. Here these results are generalized to include the magnetic dipole. Since London equations are linear, the total field of the vortex-dipole system is just the sum of the vortex field obtained in  Ref.\ \onlinecite{gcehb} with the field created by the magnetic dipole and the screening currents generated by it. From this result the total energy of the vortex-dipole system is obtained as the sum of the vortex energy in the absence of the dipole, and the vortex-dipole interaction energy  $- {\bf m}\cdot {\bf b^{vac}}$. The former,  obtained in Ref.\ \onlinecite{gcehb},   is a quadratic functional of the vortex distribution, and is written here as  the energy of interaction between the vortices in the film. The  vortex-dipole interaction energy is a linear functional of the vortex distribution, since in London theory  the field ${\bf b^{vac}}$ is  a linear in the field sources.  The functional coeficient of linearity is  interpreted as the magnetic pinning potential. The exact expression for this potential is obtained here, and its  dependence on the spatial coordinates and  on the model parameters -  magnetic moment strenght, position and orientaion, film thickness, and temperature - is studied in detail. It is also shown here that the vortex-dipole interaction energy is closely related to the screening current induced by the dipole:  the change $-{\bf m}\cdot \delta {\bf b^{vac}}$ due to  small deformation in the shape of the vortex lines is equal to the negative of the work done by Lorentz force of the screening current during the deformation.   

In the case of straight vortex lines the interaction energy $- {\bf m}\cdot {\bf b^{vac}}$, is found to be identical to that obtained by Milo\v{s}evi\'c et al \cite{myp}. Thus, with the  exception of   Ref.\ \onlinecite{coff}, the earlier results mentioned above for the energy of interaction between straight vortex lines and the magnetic dipole agree one another and with the one obtained here. 

Minimization of the vortex-dipole system energy functional is not feasible in general because  it involves infinite many degrees of freedom which are required to describe arbitrary configurations of curved vortex lines. Here the minimization  is carried out approximately using the following method. First, the exact London theory results are used to formulate a description of the vortex-dipole system  on a cubic lattice. This description preserves the physics of London theory, and has the advantage that arbitrary configurations of vortex lines can be described by a finite number of variables. Second, the  vortex-dipole system energy functional is  written in terms of these variables and minimized numerically, using simulated annealing techniques.

This paper is organized as follows. In Sec.\ \ref{sec.vmd} the exact solutions of London equations for the vortex-dipole system are obtained, and the total energy is calculated. In view of the diversity of formulas for the vortex-dipole interaction energy obtained by the earlier workers cited above, the energy calculation in Sec.\ \ref{sec.vmd} is carried out in detail. Two models for the magnetic dipole are considered: a small current loop and a permanent dipole.  In order to obtain the vortex-dipole interaction energy as $- {\bf m}\cdot {\bf b^{vac}}$ it is  fundamental to use particular properties of the solutions of London equations reported in Ref.\ \onlinecite{gcehb}. In  order to make  the contact with  Ref.\ \onlinecite{gcehb} easier,  the present paper uses the same notation.  The mathematical details of the calculations described in  Sec.\ \ref{sec.vmd} are given in Appendix\ \ref{sec.math}, and the relationship between the vortex-dipole interaction energy and the screening current is demonstrated in Appendix\ \ref{sec.wlf}.
In Sec.\ \ref{sec.vmef} the results of Ref.\ \onlinecite{gcehb} are used to write the total energy of the  vortex-magnetic dipole system  as a functional of the vortex distribution. First, in Sec.\ \ref{sec.vvi}, the vortex energy in the absence of the dipole is written in terms of vortex-vortex interactions. Then, in Sec.\ \ref{sec.pinp}, the vortex-magentic dipole interaction energy is written in terms of a magnetic pinning potential and the dependence of this potential on the spatial coordinates and parameters of the model are studied in detail.   Applications of these results of are considered in Sec.\ \ref{sec.appl}. First, the energy functional for straight vortex lines is obtained and compared with earlier results in Sec.\ \ref{sec.svl}. Then,  minimization of the vortex-dipole system energy functional is discussed in Sec.\ \ref{sec.llm}. Finally, the conclusions of this paper are stated in Sec.\ \ref{sec.conc}.


\section{vortex-magnetic dipole interaction}
\label{sec.vmd}

\begin{figure}
\centerline{\includegraphics[scale=0.25]{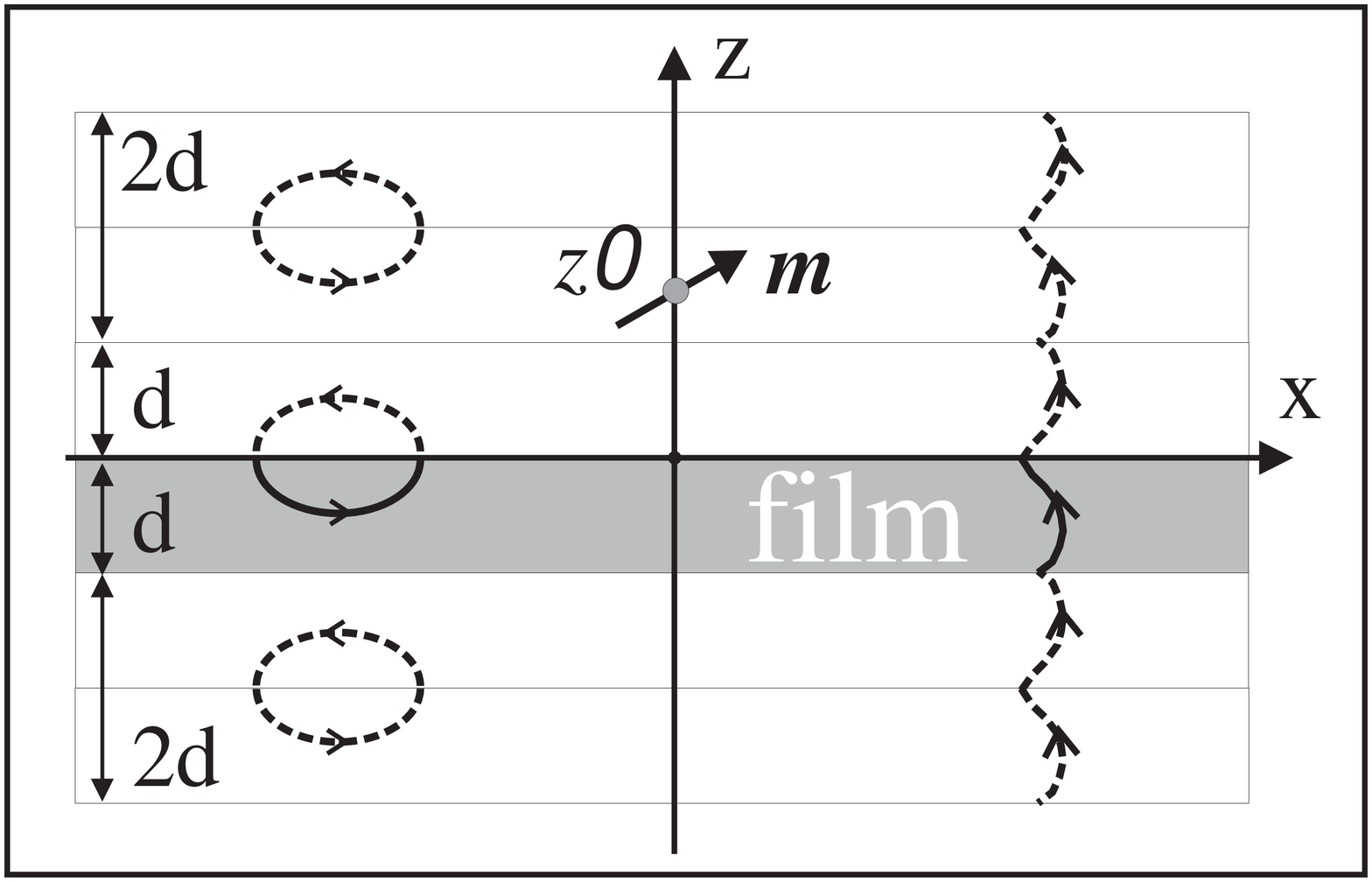}}
\caption{Superconducting film and magnetic dipole, ${\bf m}$, at $({\bf r}_{0\perp}=0,z_0)$. An example of vortex lines (full lines) and respective  images (dashed lines). }
\label{fig.fig1}
\end{figure}

The  film is assumed to be planar, with surfaces parallel to each other and to the $x-y$ plane, and of thickness $d$ occupying the region $-d\leq z \leq 0$. The superconductor is  isotropic, characterized by the penetration depth $\lambda$. The  magnetic dipole, ${\bf m}$, is placed above the film at ${\bf r}_0 =(x_0,y_0,z_0)\equiv ({\bf r}_{0\perp},z_0) $ ($z_0>0$). 
(Fig.\ \ref{fig.fig1}). 

\begin{figure}
\centerline{\includegraphics[scale=0.15]{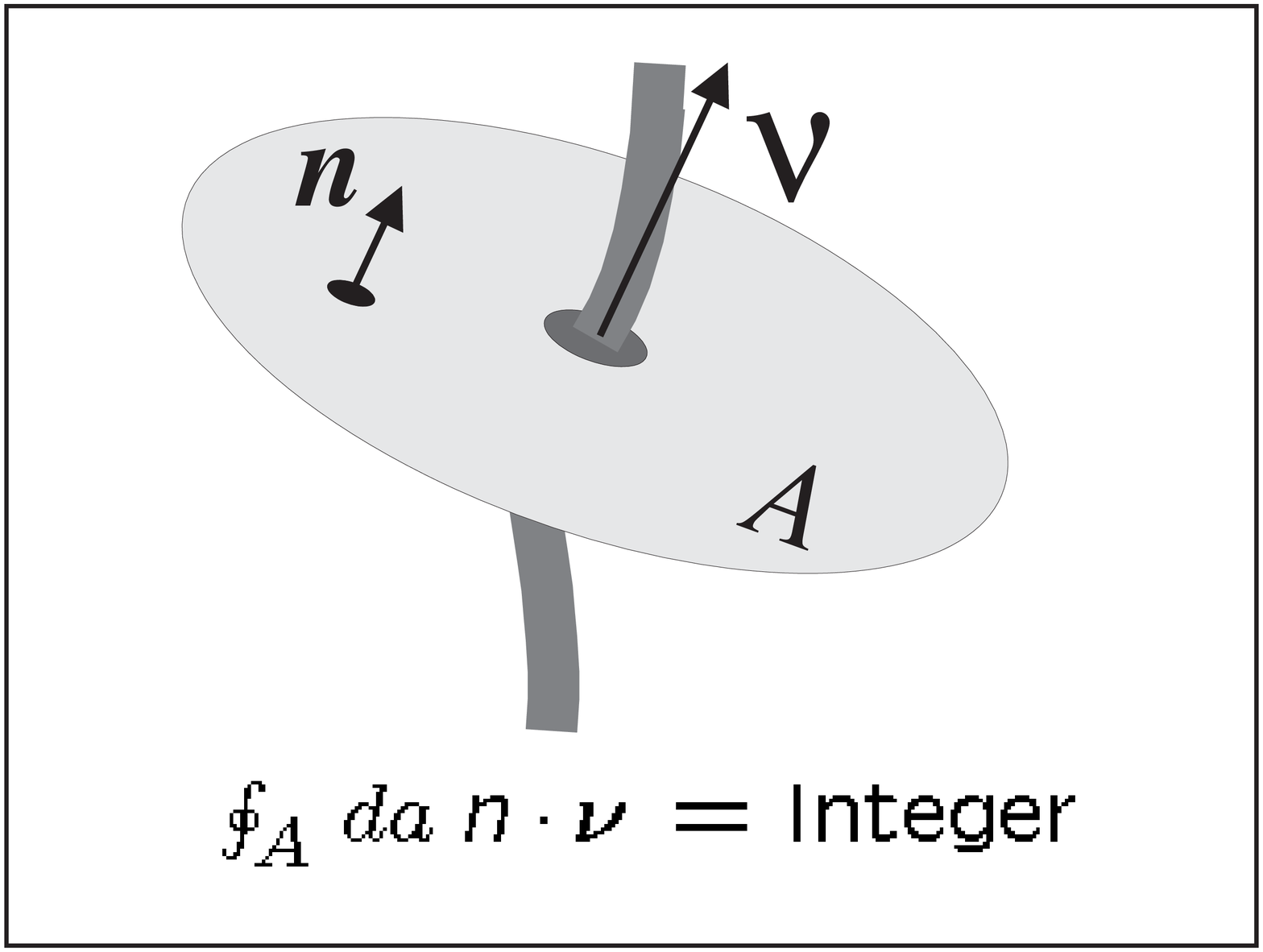}}
\caption{Physical interpretation of the vectorial vorticity distribution. The normal ${\bf n}$ is parallel to ${\mbox{\boldmath $\nu$}}({\bf r})$.  }
\label{fig.fig2}
\end{figure}

The magnetic  field of the combined vortex-dipole system is written as
  \begin{eqnarray} 
  {\bf b}^{\rm (in)} &= &{\bf b}^{\rm film} + {\bf b}^{\rm in}_{\bf m}\; \; (-d<z<0) \nonumber \\
  {\bf b}^{\rm (out)} &= &{\bf b}^{\rm vac} + {\bf b}^{\rm out}_{\bf m}\;\; (z<-d,\, z>0\,),
  \label{eq.bio}
  \end{eqnarray}
where ${\bf b}^{\rm film}$ and ${\bf b}^{\rm vac}$ are, respectively, the vortex magnetic fields inside the film and in vacuum. The fields  ${\bf b}^{\rm in}_{\bf m}$ and 
${\bf b}^{\rm out}_{\bf m}$ are  the dipole magnetic fields inside the film and in vacuum, respectively.   
The vortex fields where obtained in Ref.\ \onlinecite{gcehb} by the method of images. According to it, any configuration of vortex lines in the film is characterized by a vectorial vorticity distribution
${\mbox{\boldmath $\nu$}}({\bf r})$ or its Fourier transform
${\mbox{\boldmath $\nu$}}({\bf k})$ \cite{gms} defined as
  \begin{equation}  
  {\mbox{\boldmath $\nu$}}({\bf k})=\int\! d^3r\; e^{-i{\bf k}
  \cdot{\bf r}} {\mbox{\boldmath $\nu$}}({\bf r}) \, .
  \label{eq.nuk}
  \end{equation}
For vortex lines with vanishing core diameter,
${\mbox{\boldmath $\nu$}}({\bf k})$ is given by a sum of
line integrals along the vortices,
  \begin{equation}  
  {\mbox{\boldmath $\nu$}}({\bf k})= \sum_j \! \oint d{\bf l}_j \,
  e^{-i{\bf k}\cdot{\bf r}_j}\, .
  \label{eq.nukcl}
  \end{equation}
The	physical meaning of the vectorial vorticity distribution is that the flux of 
${\mbox{\boldmath $\nu$}}({\bf r})$ through an area perpendicular to it is an integer whose absolute value is the number of flux quanta carried by the vortex line and the sign is that of the magnetic flux through the surface, as illustrated in  Fig.\ \ref{fig.fig2}.

Inside the film, the magnetic field ${\bf b}^{\rm film}({\bf r})$
satisfies the inhomogeneous London equation
  \begin{equation}  
  -\lambda^2\nabla^2 {\bf b}^{\rm film} + {\bf b}^{\rm film}  =
  {\phi_0} {\mbox{\boldmath $\nu$}}  \,.
  \label{eq.lonfm}
  \end{equation}
Outside the film, assuming vacuum,
the magnetic field, ${\bf b}^{\rm vac}$ can be derived from a scalar potential that
satisfies the Laplace equation
  \begin{equation}  
  {\bf b}^{\rm vac}= -{\bf \nabla} \Phi \,, \; \nabla^2 \Phi=0 \,.
  \label{eq.hvacph}
  \end{equation}
The boundary conditions at the surfaces between the superconductor and the vacuum ($z=0$ and $z=-d$) are that the perpendicular component of the current vanishes and that the magnetic field is continuous. The method of images  defines a vortex distribution  in all space 
  ($-\infty < z < \infty$ ),  $ {\mbox{\boldmath ${\nu}$}}^{\rm vi}({\bf r})$, such that the current generated by it satisfies the boundary conditions,  and that it coincides with the  prescribed vorticity inside the film. In Ref.\ \onlinecite{gcehb} it is shown that   
 ${\mbox{\boldmath ${\nu}^{\rm vi}$}}$ consists of the vortex distribution  ${\mbox{\boldmath ${\nu}$}}$ and its specular images at the film surfaces. This gives a periodic vortex distribution  in the $z$-direction with period $2d$.
For the basic interval $-d \leq z \leq d\; \; \;$, ${\mbox{\boldmath ${\nu}^{\rm vi}$}}({\bf r})$ is given by 
  \begin{eqnarray}  
  {\mbox{\boldmath ${\nu}$}}^{\rm vi}({\bf r}) =&
  {\mbox{\boldmath $\nu$}}({\bf r}),~~~~~~~~  &-d\leq z \leq 0 \,,
  \nonumber \\
  {\mbox{\boldmath ${\nu}$}}^{\rm vi}_{\perp}(x,y,z) =&
  -{\mbox{\boldmath $\nu$}}_{\perp}(x,y,-z),\,~&~\,~0\leq z \leq d \,,
  \nonumber \\
  {\nu}^{\rm vi}_{z}(x,y,z)=& ~\nu_{z}(x,y,-z), &~\,~0\leq z \leq d\,,
  \label{eq.nuvi}
  \end{eqnarray}
where ${\perp}$ stands for the vector component parallel to
the $x-y$ plane. An example is shown in Fig.\ \ref{fig.fig1}. The magnetic field inside the film is then
  \begin{equation}  
  {\bf b}^{\rm film} = {\bf b}^{\rm vi} + {\bf b}^{\rm stray} \,,
  \label{eq.bfm}
  \end{equation}
where ${\bf b}^{\rm vi}$ is the field produced by the vortex
distribution and its images, and is obtained by solving London
equation, Eq.\ (\ref{eq.lonfm}), in all space with
${\mbox{\boldmath ${\nu}$}}^{\rm vi}({\bf r})$ as the field
source. The stray field inside the film, ${\bf b}^{\rm stray}$, is a
solution of the homogeneous London equation.

The magnetic field of the dipole inside the film, ${\bf b}^{\rm in}_{\bf m}$, is a solution of the homogeneous London equation. Outside the film, ${\bf b}^{\rm out}_{\bf m}$  is  the sum of the magnetic dipole field in the absence of the superconductor and the field of the screening current induced by the dipole.  The boundary conditions are the same as those for the vortex fields. The  magnetic field of the dipole is discussed in detail in 
 Ref.\ \onlinecite{myp}

The total energy of the  vortex-dipole system, $E_T$, defined as the sum of the kinetic energy of the supercurrent in the film with  magnetic energy inside and outside the film,  can   be written as
  \begin{equation} 
  E_T=E_{\rm in} + E_{\rm out} \,,
  \label{eq.etot}
  \end{equation}
where $E_{\rm in}$ is the London  energy of the supercurrent and of the magnetic field inside the film, and $E_{\rm out}$ is the  energy of the vacuum magnetic field
  \begin{eqnarray}  
  &&E_{\rm in} = \int\!\frac{d^2r_{\perp}}{8\pi} 
  \int^0_{-d}\!\! dz \, [\lambda^2\mid{\bf \nabla}\times {\bf b}^{\rm (in)}\mid^2 + \mid {\bf b^{\rm (in)}}\mid^2]  \,,
  \nonumber \\
  &&E_{\rm (out)}=\int\frac{d^2r_{\perp}}{8\pi} \Big[ \int^{\infty}_0\! dz\,  + \int^{-d}_{-\infty}\!\! dz\, \Big]\, \mid{\bf b}^{\rm (out)}\mid^2 .
  \label{eq.enfenv}
  \end{eqnarray}

Now the coupling between the magnetic dipole and the superconductor is considered for two  particular models for the  dipole:  a small current loop, and a point dipole. 

In the case where the magnetic dipole is a small current loop, the change in the total energy  resulting from a small change in the vortex distribution, $\delta E_T$, must be equal to the work done by the external source attached to the loop to keep the current constant during the 
change, $\delta E_{ext}$, that is 
\begin{equation}
\delta E_T-\delta E_{ext}=0
\label{eq.de1}
\end{equation}
 where
\begin{equation}
\delta E_{ext} = \frac{1}{c}\int\, d^3r\, {\bf j}_{ext}\cdot \delta {\bf a^{vac}} \; ,
\label{eq.dex1}
\end{equation}
where ${\bf j}_{ext}$ is the current density in the loop, and $\delta {\bf a^{vac}}$, defined by 
$\delta{\bf b^{vac}}=  {\mbox{\boldmath ${\nabla}$}}\times \delta{\bf a^{vac}}$, is the change in the vector potential of the ${\bf b^{vac}}$ field. Expanding $\delta{\bf a^{vac}}$ around ${\bf r}_0$ as 
\[\delta{\bf a^{vac}}({\bf r})=\delta{\bf a^{vac}}({\bf r}_0) + 
({\bf r}-{\bf r}_0)\cdot{\mbox{\boldmath ${\nabla}$}} \delta{\bf a^{vac}}({\bf r}_0)\]
it is straightforward to show that
\begin{equation}
\delta E_{ext} = {\bf m}\cdot \delta {\bf b^{vac}}({\bf r}_0) \; .
\label{eq.deext2}
\end{equation}
It turns out that this is the only vortex-dipole interaction term in Eq.\ (\ref{eq.de1}). Other possible contributions to the vortex-dipole interaction would result from cross terms in $\delta E_T$ containing  products of the vortex and dipole fields.  It is shown in Appendix\ \ref{sec.math} that these terms vanish, so that $\delta E_T$ is the same as in the absence of the  magnetic dipole. According to these results $\delta E_{vm}\equiv - {\bf m}\cdot \delta {\bf b^{vac}}({\bf r}_0)$ can be interpreted as the change in the  energy of interaction between the vortices and the dipole, 
\begin{equation}
E_{vm}= - {\bf m}\cdot {\bf b^{vac}}({\bf r}_0)\;. 
\label{eq.evm}
\end{equation}

The  same vortex-magnetic dipole interaction energy is obtained if the magnetic dipole is modeled by a permanent point dipole. This model explores the well known analogy between magneto statics and electrostatics in current free regions of space \cite{jack}. According to it,  
${\bf b}^{\rm out}_{\bf m}$ can be derived from a scalar potential, 
${\bf b}^{\rm out}_{\bf m}=-{\mbox{\boldmath $\nabla$}}\Phi_{\bf m}$  that satisfies the Poisson equation
  \begin{equation}  
  \nabla^2\,\Phi_{\bf m}= -4\pi\, {\bf m}\cdot
{\mbox{\boldmath $\nabla$}} \delta({\bf r}-{\bf r}_0)
  \label{eq.pseq}
  \end{equation}
In this case the problem is identical to that of an electric dipole in an external field. The   vortex-magnetic dipole interaction energy comes from the crossed term in $E_{\rm out}$ with   ${\bf b}^{\rm out}_{\bf m} \cdot{\bf b^{vac}}$, as shown in  Appendix\ \ref{sec.math} . This approach is the same as that used in  Refs.\ \onlinecite{wei,sah,myp}. In this case the vortex-magnetic dipole interaction energy is the work done to bring the dipole from infinity to its final position.

To summarize, the total energy of the vortex-magnetic dipole system can be written as  
\begin{equation}
E_T=E_v - {\bf m}\cdot{\bf b^{vac}}({\bf r}_0) -\frac{1}{2}{\bf m}\cdot 
{\bf b}'_{\bf m}({\bf r}_0)\; ,
\label{eq.evds}
\end{equation}
where $E_v$ is the energy of the vortex distribution alone. The last term in Eq.\ (\ref{eq.evds}) is the energy of the dipole alone in the presence of the superconductor, ${\bf b}'_{\bf m}({\bf r}_0)$
being the field of the dipole screening current at the dipole position. In this paper this term is   a constant, since ${\bf m}$ is not allowed to change. From here on this term is dropped.

The vortex-dipole interaction energy, Eq.\ (\ref{eq.evm}), can be generalized to any  distribution of permanent magnetic dipoles placed outside the film and described by the magnetization ${\bf M}$. The result is  
  \begin{equation}  
 E_{vM}= - \int\, d^3r \, {\bf M}({\bf r})\cdot\, {\bf b^{vac}}({\bf r}) \,.
 \label{eq.enfilm}
  \end{equation}
This expression is in agreement with  a general formula of classical electrodynamics  that expresses the interaction energy of a magnet in an external field in terms of the field that would exist in the absence  the magnet \cite{ll}. Here this field is  ${\bf b^{vac}}({\bf r}) $.   

\section{vortex-magnetic dipole energy functional}
\label{sec.vmef}
Here the results of Ref.\ \onlinecite{gcehb} are used to express the total energy of the vortex-dipole system as a functional of the vortex vectorial distribution.


\subsection{vortex-vortex interactions}
\label{sec.vvi}

It follows from Eqs.(32)-(35) and (22)-(28) of Ref.\ \onlinecite{gcehb} that the vortex energy, $E_{v}$, is a quadratic functional of the vectorial vortex distribution,  
which  can be written as 
  \begin{eqnarray}  
&&E_v/(\epsilon_0 \lambda) =
\int\! d^2r_{\perp}\;\int\! d^2r'_{\perp}  \int^0_{-d}\; \frac{dz}{\lambda} \;\int^0_{-d}\; 
\frac{dz'}{\lambda} \nonumber \\
&&[ {\cal G}_{\perp}(\mid{\bf r}_{\perp}- {\bf r}'_{\perp}\mid;z,z')\,
{\mbox{\boldmath $\nu$}}_{\perp}({\bf r}_{\perp},z)\cdot {\mbox{\boldmath $\nu$}}_{\perp}({\bf r}'_{\perp},z') + \nonumber \\ 
&&{\cal G}_{zs}(\mid {\bf r}_{\perp}- {\bf r}_{\perp}\mid;z,z')
 \, \nu_z({\bf r}_{\perp},z)\nu_z({\bf r}'_{\perp},z')] \;,
  \label{eq.evnu}
  \end{eqnarray}
where $\epsilon_0=(\phi_0/(4\pi\lambda))^2$ is the basic scale for energy/length. The dimensionless functions ${\cal G}_{\perp}$ and ${\cal G}_{zs}$ describe, respectively, the interactions between the components of the vorticity perpendicular and parallel  to the $z$-axis: ${\cal G}_{\perp}$ comes  from vortex-vortex and  vortex-image interactions,  whereas ${\cal G}_{zs}$ has one contribution from  vortex-vortex and vortex-image interactions, denoted ${\cal G}_{z}$, and another resulting from the energy of the stray and vacuum fields, denoted ${\cal G}_{sv}$, that is 
  \begin{equation}  
{\cal G}_{zs}( r_{\perp};z,z')= 
{\cal G}_{z}(r_{\perp};z,z')+ {\cal G}_{sv}( r_{\perp};z,z')
 \label{eq.gzs}
  \end{equation}
The functions  ${\cal G}_{\perp}$ and ${\cal G}_{z}$ are given by
\begin{eqnarray}
&&{\cal G}_{z({\perp})}({\bf r}_{\perp};z,z') = 
\pi \lambda \displaystyle \int\,\frac{d^2k_{\perp}}{(2\pi)^2}  \frac{e^{i{\bf k}_{\perp}\cdot{\bf r}_{\perp}}}{\tau} [e^{-\tau\mid z-z'\mid }+\nonumber \\ 
&& \displaystyle \frac{e^{-\tau d }\cosh{\tau(z-z')}}{\sinh{\tau d}} 
 +(-) \frac{\cosh{\tau(z+z'+d)}}{\sinh{\tau d}}] \; ,
\label{eq.gzpk}
\end{eqnarray}
where $\tau =\sqrt{k^2_{\perp}+\lambda^{-2}}$, and the plus (minus) sign is
 for ${\cal G}_{z}$ ( ${\cal G}_{\perp}$). The first terms in the brackets in Eqs.\ (\ref{eq.gzpk}) come from  bulk vortex-vortex interactions,  whereas the second and third terms come from the interactions between the vortices and their images.  These two functions need a short range cutoff to avoid unphysical divergencies  at the vortex core \cite{ehbrev}. 
The interaction function resulting from the stray and vacuum   fields is given by
\begin{eqnarray}
&&{\cal G}_{sv}({\bf r}_{\perp};z,z')=-
\displaystyle \frac{2\pi}{\lambda}\int \, \frac{d^2k_{\perp}}{(2\pi)^2} 
\; \frac{e^{i{\bf k}_{\perp}\cdot{\bf r}_{\perp}}}{k_{\perp}\tau (\sinh{\tau d})^2}  \{f_1\times\nonumber \\
&&\displaystyle
[\cosh{\tau(z+d)}\cosh{\tau(z'+d)}+ \cosh{\tau z}\cosh{\tau z'}]+  f_2\times  \nonumber\\
&&\displaystyle
[\cosh{\tau(z+d)}\cosh{\tau z'}+ \cosh{\tau z}\cosh{\tau(z'+d)}]\} \; ,
\label{eq.gsvk}
\end{eqnarray}
 where
\begin{eqnarray}
f_1&=&\displaystyle\frac{(k_{\perp}+\tau )e^{\tau d } +(k_{\perp}-\tau)e^{-\tau d }}{{\cal C}} \nonumber \\
f_2&=&\displaystyle-\frac{2k_{\perp}}{{\cal C}} \nonumber \\ 
{\cal C}&  = &(k_{\perp} \!-\! \tau)^2\,e^{-\tau d}
  - (k_{\perp} \!+\! \tau)^2\,e^{\tau d} \;.
\end{eqnarray}
The functions  
${\cal G}_{\perp}$, ${\cal G}_z$ and ${\cal G}_{sv}$ are invariant with respect to the transformations $z \longleftrightarrow z'$, and   $(z,\, z')\,\rightarrow (-z-d,-z'-d)$. They  depend  only on $d/\lambda$ and on 
$\xi/\lambda$, where $\xi$ is the vortex core radius. For thin films ($d\ll \lambda$) the term in  ${\cal G}_{\perp}$ is absent in Eq.\ (\ref{eq.evnu}), and  ${\cal G}_{zs}$ reduces to Pearl's result  \cite{vvlr}. 


\subsection{magnetic pinning potential}
\label{sec.pinp}
The vortex-dipole interaction energy is a linear functional of the vectorial vortex distribution, which  can be written as 
  \begin{equation}  
E_{vm}/(\epsilon_0\lambda)=\int\! d^2r_{\perp} \int^0_{-d}\;\frac{dz}{\lambda}\;
\nu_z({\bf r}_{\perp},z)
U_{vm}({\bf r}_{\perp},z)\;,
 \label{eq.evm2}
  \end{equation}
where $U_{vm}$ is the magnetic pinning potential. This result follows from Eq.\ (\ref{eq.evm}), and from the linear dependence of the scalar potential for the vacuum field, $\Phi$, Eq.\ (\ref{eq.hvacph}), on the z-component of the vorticity obtained in Ref.\ \onlinecite{gcehb}. Writing $\Phi$  as 
  \begin{eqnarray}  
\Phi({\bf r}_{\perp},z)&=&\int\! d^2r'_{\perp} \int^0_{-d}dz'\,
\nu_z({\bf r'}_{\perp},z')\times \nonumber \\
& &{\cal K}(\mid{\bf r}_{\perp}-{\bf r}'_{\perp}\mid;z,z') \;,
 \label{eq.phkp}
  \end{eqnarray}
it follows from  Eq.\ (\ref{eq.evm}) that the magnetic pinning potential is given by 
  \begin{equation}  
U_{vm}({\bf r}_{\perp},z)= \frac{{\bf m}}{\phi_0\lambda}\cdot
 {\mbox{\boldmath $\nabla$}}_0\, {\cal K}(\mid{\bf r}_{0\perp}-{\bf r}_{\perp}\mid;z_0,z)\; 
 \label{eq.uvm}
  \end{equation}
The expression for the kernel ${\cal K}$ follows from Eqs.(23), (25), (27),  and (20) of Ref.\ \onlinecite{gcehb}. The result is 
\begin{eqnarray}
&&{\cal K}(\mid {\bf r}\mid;z_0,z)= -(4\pi)^2\lambda \displaystyle\int\,\frac{d^2k_{\perp}}{(2\pi)^2}
\frac{e^{i{\bf k}_{\perp}\cdot {\bf r}} e^{-k_{\perp}z_0}}
{k_{\perp}} \times \nonumber \\
&& \displaystyle \frac{(k_{\perp}+\tau)e^{\tau(z+d)} - (k_{\perp}-\tau)e^{-\tau(z+d)}}
{{\cal C}}\; .
\label{eq.kapk} 
\end{eqnarray}
According to Eq.\ (\ref{eq.evm2}) $(\epsilon_0\lambda)U_{vm}$ is the energy of interaction  of a vortex element parallel to the $z$-direction  with the dipole. Note that the vortex-dipole interaction energy, Eq.\ (\ref{eq.evm2}), does not depend  on the component of the vorticity perpendicular to the $z$-direction, ${\mbox{\boldmath $\nu$}}_{\perp}$. However, $\nu_z$ and ${\mbox{\boldmath $\nu$}}_{\perp}$ are not independent, since vortex lines form closed loops or lines  that begin and terminate at the film surfaces (${\bf \nabla}\cdot{\mbox{\boldmath $\nu$}}=0$). 
The magnetic pinning potential, $U_{vm}$, depends only on the scaled variables  $d/\lambda$, $z_0/\lambda$ and  ${\bf m}/\phi_0\lambda$.  Since $\lambda$ depends on the temperature, $U_{vm}$ is temperature dependent, as shown next. 

\begin{figure}
\centerline{\includegraphics[scale=0.40]{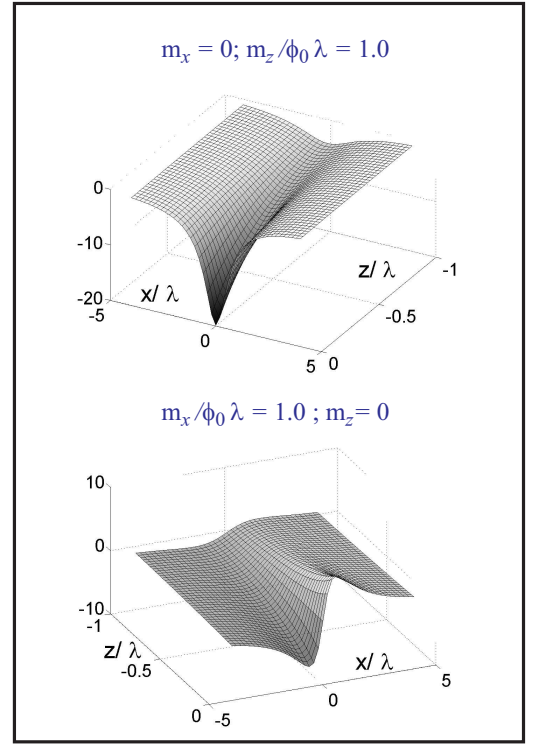}}
\caption{Dependence of the magnetic pinning potential in the plane of the dipole on the coordinates $x$ and $z$ for a film with $d=\lambda$. }
\label{fig.fig3}
\end{figure}

\begin{figure}
\centerline{\includegraphics[scale=0.33]{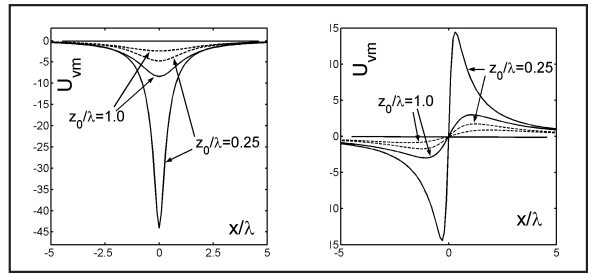}}
\caption{Dependence of the magnetic pinning potential in the plane of the dipole for a film with $d=\lambda$  on the coordinate $x$ at $z=0$ (full line) and $z=-d$ (dashed line) for two values of  the dipole height $z_0$. Left panel: $m_x=0$, $m_z=\phi_0\lambda$. Right panel: $m_x=\phi_0\lambda$, $m_z=0$.  }
\label{fig.fig4}
\end{figure}

\begin{figure}
\centerline{\includegraphics[scale=0.33]{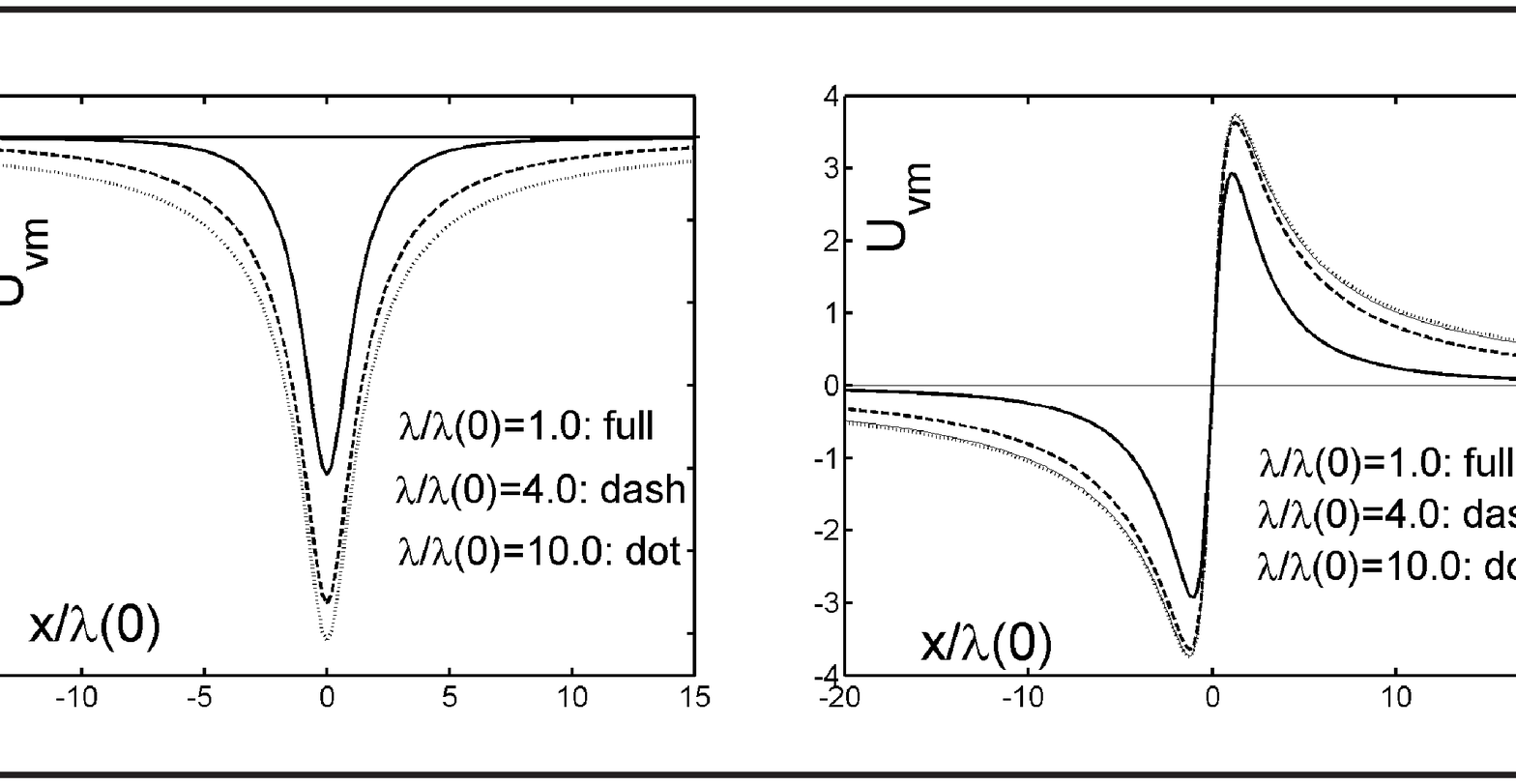}}
\caption{Dependence of the magnetic pinning potential in the plane of the dipole, and at $z=0$, on the coordinate $x$ for a film with $d=2\lambda$ at different temperatures,  defined by $\lambda/\lambda(0)$. Dipole height $z_0=\lambda$. Left panel: $m_x=0$, $m_z=\phi_0\lambda$. Right panel: $m_x=\phi_0\lambda$, $m_z=0$. }
\label{fig.fig5}
\end{figure}

Assuming that the dipole is located in the $x-z$ plane and that ${\bf r}_{0\perp}=0$, the pinning potential can be written as
  \begin{eqnarray}  
&U_{vm}({\bf r}_{\perp},z)=
 \displaystyle \frac{ m_z}{\phi_0\lambda}\frac{\partial {\cal K}( r_{\perp};\,z_0,\,z) }{\partial z_0} - \nonumber \\
& \displaystyle \frac{ m_x}{\phi_0\lambda}\cos{\varphi}\,\frac{\partial {\cal K}( r_{\perp};\,z_0,\,z)}{\partial r_{\perp}}
\; ,
 \label{eq.uvmb}
  \end{eqnarray}
where $\varphi$ is the angle between ${\bf r}_{\perp}$ and the $x$-axis.
Simple analytical expressions $U_{vm}$ exist in two limiting situations: large distances and thin films. The behavior of  at large distances follows from the $k_{\perp}\rightarrow 0$ limit in Eq.\ (\ref{eq.kapk}). The result is 
  \begin{eqnarray}  
&U_{vm}({\bf r}_{\perp},z)= -8\pi \displaystyle \frac{\cosh{((z+d)/\lambda)}}
{\sinh{(d/\lambda)}}\times \nonumber \\
& \displaystyle \{\frac{ m_z}{\phi_0\lambda}\frac{z_0\lambda^2}{(r^2_{\perp}+z^2_0)^{3/2}}- 
\frac{ m_x}{\phi_0\lambda}\cos{\varphi}\frac{r_{\perp}\lambda^2}{(r^2_{\perp}+z^2_0)^{3/2}}\} 
\; .
 \label{eq.uvmld}
  \end{eqnarray}
This expressions is valid  for  $\sqrt{(r^2_{\perp}+z^2_0)}\gg \lambda$ in films that are not too thin, and in thin films ($d\ll \lambda$) for $\sqrt{(r^2_{\perp}+z^2_0)}\gg \Lambda=2\lambda^2/d$.
In thin films  for $\sqrt{(r^2_{\perp}+z^2_0)}\ll \Lambda$, 
$U_{vm}$ is given by
\begin{eqnarray}
&U_{vm}({\bf r}_{\perp},z)= -4\pi  \displaystyle \{
\{\frac{ m_z}{\phi_0}\frac{1}{(r^2_{\perp}+z^2_0)^{1/2}}-
\nonumber \\ 
& \displaystyle \frac{ m_x}{\phi_0}\frac{\cos{\varphi}}{r_{\perp}}[1-
\frac{z_0}{(r^2_{\perp}+z^2_0)^{1/2}}]\} 
\; .
\label{eq.uvmtf}
\end{eqnarray}
Note that for thin films $U_{vm}$, Eq.\ (\ref{eq.uvmtf}), is independent of the temperature, since $\lambda$ drops out. 
For films of finite thickness, and at short distances,  
$U_{vm}$ has to be calculated numerically. Some results for  $U_{vm}$ on the plane of the dipole  are shown in Figs.\ \ref{fig.fig3}, and \ref{fig.fig4}.  For ${\bf m}$ perpendicular to the film surfaces ($m_x=0$), $U_{vm}$ is invariant under rotations around the $z$-axis, so that the graphs in Figs.\ \ref{fig.fig4} and \ref{fig.fig5}, left panels, represent the dependence of $U_{vm}$ on $r_{\perp}$. For ${\bf m}$ parallel to the film surfaces ($m_z=0$), the dependence of  $U_{vm}$ on $r_{\perp}$ in a plane rotated by $\varphi$ around the $z$-axis with respect to  the $x-z$ plane is just that shown in Figs.\ \ref{fig.fig4}and \ref{fig.fig5}, right panels, multiplied by $\cos{\varphi}$ (Eq.\ (\ref{eq.uvmb})). These results show that  the magnetic pinning potential penetrates a distance $\sim \lambda$ into the film, and that its range parallel to the film surfaces is  a few $\lambda$.  The temperature dependence of $U_{vm}$, shown in Fig.\ \ref{fig.fig5}, is that increasing the temperature increases both the range and the absolute value of $U_{vm}$. 
%
%
%
%

\section{applications}
\label{sec.appl}

In this section the results derived above are applied to study the equilibrium states of vortices induced by the magnetic dipole. 
To obtain these states   in the absence of an applied  magnetic field  it is necessary to minimize the energy $E_T$, Eq.\ (\ref{eq.evds}), with respect to the vorticity distribution ${\mbox{\boldmath $\nu$}}$. The equilibrium state, in the absence of an externally applied field, will contain vortices if there is a configuration of vortex lines for which 
$E_{T}=E_v+E_{vm}$ is a minimum and negative. Minima with $E_T>0$  are metastable, since for the film without vortices $E_v=E_{vm}=0$. Minima with  $E_T<0$ can occur if there are configurations of vortex lines for which $E _{vm}$ is sufficiently negative to overcome the positive definite vortex-vortex interaction energy $E_v$. The shapes of the vortex lines generated by the dipole are dictated by the competition between $E_v$ and $E _{vm}$. The spatial inhomogeneity of the magnetic pinning potential also plays an important role in determining the shapes, as shown in Sec.\ \ref{sec.llm}. 

According to the discussion in Sec.\ \ref{sec.vmef},  the equilibrium states depend only on the scaled parameters  $d/\lambda$, $z_0/\lambda$, and ${\bf m}/\phi_0\lambda$, and on the temperature. If thermal vortex fluctuations are neglected, which is appropriate for low-T$_c$ superconductors, the temperature dependence comes only from $\lambda$. For a given film and magnetic dipole  $d$, $z_0$ and ${\bf m}$ are fixed, but the parameters  $d/\lambda$, $z_0/\lambda$ and  ${\bf m}/\phi_0\lambda$ change with temperature, leading to nontrivial changes in the equilibrium state, as  discussed in Sec.\ \ref{sec.llm}.

The problem of minimizing $E_T$  with respect to the vortex distribution is, in general, a formidable task. The simplest case is that of straight vortex lines perpendicular to the film surfaces, where the functional dependence of $E_T$ on the vortex distribution can be described by a finite number of degrees of freedom, as discussed in Sec.\ \ref{sec.svl}. For films of finite thickness the vortex lines are not in general straight, and $E_T$ depends on infinite many degrees of freedom required to describe arbitrary configurations of curved vortex lines.
The minimization of $E_T$ can only be carried approximately, by reducing the degrees of freedom to a discrete set. A method to do this is discussed in Sec.\ \ref{sec.llm}

\subsection{straight vortex lines}
\label{sec.svl}

Here it is assumed here that the vortex lines in the film are straight and perpendicular to the film surfaces.  In this case ${\mbox{\boldmath $\nu$}}_{\perp}=0$ and $\nu_{z}$ is given by
\begin{equation}
\nu_{z}({\bf r}_{\perp},z)=\sum_{(i)}\, 
q_i\,\delta({\bf r}_{\perp}-{\bf R}_i)\; ,
\label{eq.}
\end{equation} 
Where  $q_i=0,\,\pm1, \pm 2..$ is the vorticity of the $i$-th vortex line   and ${\bf R}_i$ its position in the $x-y$ plane.
The  energy is  then  
 \begin{eqnarray}
&E_T/(\epsilon_0\lambda)= \{\sum_{(i,j)}\, q_i\,q_j\,
 U_{vv}(\mid {\bf R}_i -{\bf R}_j\mid) + \nonumber \\ 
&\sum_{(i)}\, q_i \,U^{\rm line}_{vm}\,({\bf R}_i)\}\; 
\label{eq.esvl}
\end{eqnarray}
where 
\begin{eqnarray}
&U_{vv}(\mid {\bf R}_i -{\bf R}_j\mid)= \nonumber \\
&\displaystyle \int^0_{-d}\; \frac{dz}{\lambda} \;\int^0_{-d}\; 
\frac{dz'}{\lambda} {\cal G}_{zs}(\mid {\bf R}_i -{\bf R}_j\mid;z,z')\;,
\label{eq.uvv}
\end{eqnarray}
and
\begin{equation}
U^{\rm line}_{vm}({\bf R}_i)=\int^0_{-d}\; \frac{dz}{\lambda} U_{vm}({\bf R}_i,z)\;.
\label{eq.uvml}
\end{equation}
The interaction energy  of a vortex line pair (${\bf R}_i\neq {\bf R}_j$)is 
$(\epsilon_0\lambda)\, 2U_{vv}(\mid {\bf R}_i -{\bf R}_j\mid)$, and $(\epsilon_0\lambda)\, U_{vv}(0)$ is the vortex line self-energy. The vortex-vortex interaction energy $U_{vv}$ is discussed in detail in Ref.\ \onlinecite{gcehb}.

The interaction energy   of the vortex line with the magnetic dipole is $(\epsilon_0\lambda)\, U^{\rm line}_{vm}({\bf R}_i)$.
The expression for $U^{\rm line}_{vm}$, is obtained from Eq.\ (\ref{eq.uvm}), (\ref{eq.kapk}), and (\ref{eq.uvml}) as 
\begin{eqnarray}  
&&U^{\rm line}_{vm}({\bf R})= \displaystyle-(4\pi)^2 \int\,\frac{d^2k_{\perp}}{(2\pi)^2}\nonumber \\
&&\displaystyle (\frac{ m_x}{\phi_0\lambda}ik_x-\frac{m_z}{\phi_0\lambda}k_{\perp}) e^{-i{\bf k}_{\perp}\cdot {\bf R}} e^{-k_{\perp}z_0}\times \nonumber \\
&&\displaystyle \frac{(k_{\perp}+\tau)e^{\tau d}+ (k_{\perp}-\tau)e^{-\tau d}-2k_{\perp}}
{{\cal C}k_{\perp}\tau} \;. 
  \label{eq.uvmlb}
  \end{eqnarray}
It is straightforward to shown that this expression is identical to that obtained in Ref.\ \onlinecite{myp}. 
It can  be shown, using the results of Appendix\ \ref{sec.wlf}, that $U^{\rm line}_{vm}({\bf R})$ can also be written as the negative of the work done  by the Lorentz force of the dipole screening current  to bring the vortex line from a position far from the dipole to ${\bf R}$. In Ref.\ \onlinecite{myp} the vortex-dipole interaction energy is written as the sum of minus  one-half of the Lorentz force work with  $- {\bf m}\cdot {\bf b^{vac}}/2$. 

The total energy Eq.\ (\ref{eq.esvl}) is a functional of the vorticities ($q_i$) and positions (${\bf R}_i$) of the vortex lines. In order to obtain the equilibrium vortex configurations $E_T$ has to be minimized with respect to these variables. The minimization involves only a finite number of variables, and can be carried out numerically with modest computational resources. This minimization will be discussed elsewhere.

The equilibrium vortex configurations created by the magnetic dipole will consist of  straight vortex lines  only for thin films ($d\ll\lambda$).  For thick films $d\gg \lambda$, the self energy and the  vortex-vortex interaction energy grow with the thickness of the film, $d$, whereas the  vortex-dipole interaction energy does not, since $U_{vm}$ is limited to region of depth $\lambda$ from the film surface closest to the dipole.

\subsection{lattice London model }
\label{sec.llm}

In this section an approximate method to obtain  equilibrium vortex configurations induced by the magnetic dipole is presented. It is based on a discretization of the vortex degrees of freedom  called lattice London model. This model was introduced several years ago to study vortex fluctuations in high-T$_c$ superconductors \cite{gmcllm}. In the present context the lattice London model is useful because it requires only a  finite number of degrees of freedom to describe curved vortex line configurations, and because it preserves the essential physical ingredients  of the  vortex-dipole system in the London limit. 

\begin{figure}
\centerline{\includegraphics[scale=0.25]{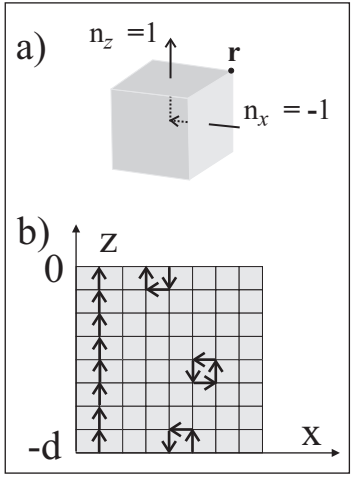}}
\caption{a) Graphical representation of integer vorticities at lattice point ${\bf r}$. 
b )Vortex loops and vortex lines used in the numerical minimization method to generate vortex configurations. }
\label{fig.fig6}
\end{figure}

The  vortex distribution in the lattice London model is represented by  integer variable placed on three-dimensional mesh with cubic unit cell of side $a\sim \xi$, and subjected to periodic boundary conditions. At each lattice site there are three integers $n_{\mu}=0,\pm 1,\pm 2...$, one for each spatial direction $\mu =x,y,z$. From the  configuration of the variables $n_{\mu}$ at each lattice site, the configuration of vortex lines follow by associating arrows with the $n_{\mu}$, as shown in Fig.\ \ref{fig.fig6}.a. Essentially, the lattice London model restricts the vorticity ${\mbox{\boldmath $\nu$}}$ to point in one of the $x,y,z$ directions, and  $n_{\mu}$ represents the flux of ${\mbox{\boldmath $\nu$}}$ through the face of the cubic unit cell  perpendicular to the $\mu$ direction (Fig.\ \ref{fig.fig6}.a). For  bulk  and semi-infinite superconductors the lattice London model is an exact discretization of London theory on a cubic lattice, as shown in Ref.\ \onlinecite{gmcllm}.   
For the problem under consideration here, the lattice London model is   an approximation. It consists in replacing the film by a cubic mesh of lattice constant $a\sim \xi$, where the vortices are defined as discussed above. The cubic mesh is subjected to periodic boundary conditions in the $x$ and $y$ directions. For the sake of simplicity, in what follows it is assumed  that the vortex lines generated by the dipole are in the plane of the dipole. This reduces the search for vortex line configurations to two-dimensions ($x-z$ plane), so that  $n_y=0$. This assumption is justified for the equilibrium vortex configurations discussed  here.  The vortex-dipole system energy functional is obtained from the one derived above for the continuum London model as described next.
The vortex-vortex interaction energy is taken as  
\begin{eqnarray} 
 &E_v/(\epsilon_0 \lambda) =(\frac{a}{\lambda})^2\sum_{i,j}\nonumber \\
&\{ n_x(x_i,z_i)n_x(x_j,z_j)\Gamma_{\perp}(\mid x_i- x_j\mid;z_i,z_j) \nonumber \\
&  n_z(x_i,z_i)n_z(x_j,z_j)\Gamma_{zs}(\mid x_i- x_j\mid;z_i,z_j) \}
 \;,
  \label{eq.evllm}
  \end{eqnarray} 
and the vortex-dipole interaction energy as,
\begin{equation}
E_{vm}/(\epsilon_0\lambda)=\frac{a}{\lambda}\sum_i \;
n_z(x_i,z_i)\Upsilon_{vm}(x_i,z_i)\;.
 \label{eq.evmllm}
  \end{equation}
The  functions 
$\Gamma_{\perp}$ and $\Gamma_{zs}$ and $\Upsilon_{vm}$ are obtained from ${\cal G}_{\perp}$,  ${\cal G}_{sv}$, and $U_{vm}$, respectively, by making the latter ones periodic in the lattice along the $x$ and $y$ directions. This consists in  substituting ${\bf k}_{\perp}= (k_x,\,k_y)$ in Eqs.\ (\ref{eq.gzpk}), (\ref{eq.gsvk}), and (\ref{eq.kapk}) by  ${\mbox{\boldmath $\kappa$}}=[2a^{-1}\sin{(k_xa/2)},\,2a^{-1}\sin{(k_ya/2)}]$, and by replacing the integrals over ${\bf k}_{\perp}$ in the same equations by sums over the reciprocal lattice of the cubic mesh.
This procedure is justified by the exact results of Ref.\ \onlinecite{gmcllm}.  
\begin{figure}
\centerline{\includegraphics[scale=0.35]{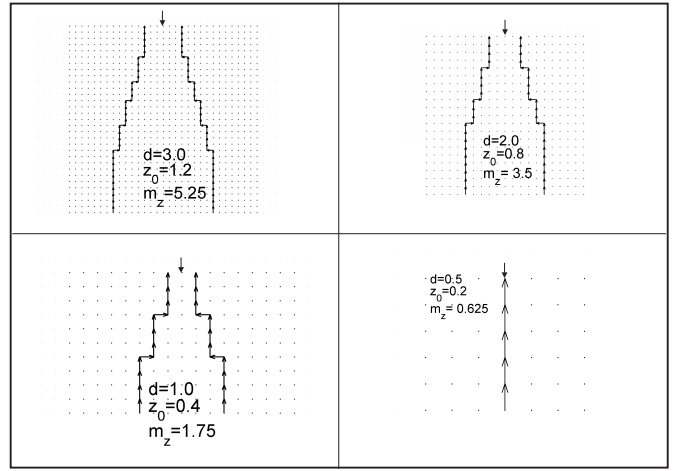}}
\caption{Equilibrium vortex configurations generated by a magnetic dipole perpendicular to the film surfaces, and located at $z_0\lambda$ from the point indicated by the arrow. $d$  in units of $\lambda$, $m_z$ in units of $\phi_0 \lambda$. Mesh spacing $a=0.1\lambda$.}
\label{fig.fig7}
\end{figure}
 
\begin{figure}
\centerline{\includegraphics[scale=0.35]{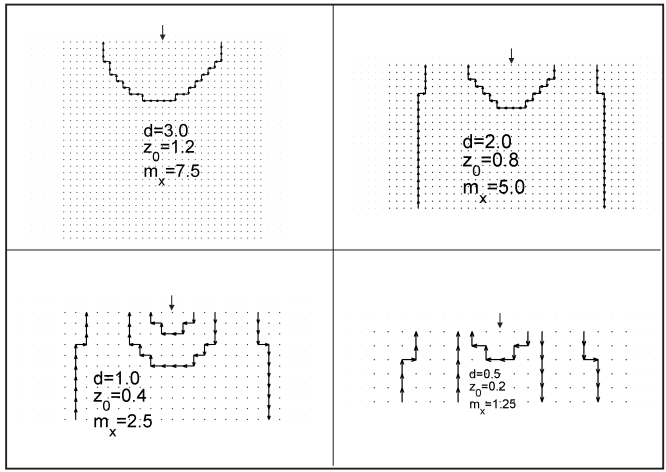}}
\caption{Equilibrium vortex configurations generated by a magnetic dipole parallel to the film surfaces, and located at $z_0\lambda$ from the point indicated by the arrow. $d$  in units of $\lambda$, $m_x$ in units of $\phi_0 \lambda$. Mesh spacing $a=0.1\lambda$.}
\label{fig.fig8}
\end{figure}
To minimize the functional $E_T=E_v + E_{vm}$, Eqs.\ (\ref{eq.evllm}) and (\ref{eq.evmllm}), with respect to $n_x$ and $n_z$  simulated annealing is used ,together with the following  procedure to generate the vortex line configurations \cite{gmcllm}. First, it is attempted to add vortex loops at every lattice site: square loops at sites not on the film surfaces and open loops at surface sites. Second, it attempted to add straight vortex lines, perpendicular to the film surfaces at every position $x_i$. This is illustrated in Fig.\ \ref{fig.fig6}.b.

The numerical minimization of $E_T$ is carried out for a few parameter values with $a/\lambda=0.1$. The results are shown in Figs.\ \ref{fig.fig7} and \ref{fig.fig8}.  These figures can also be viewed as representing the evolution of the equilibrium vortex configurations for the  film with $d=3.0\lambda(0)$, at $T=0$, with increasing temperature. The values of the scaled parameters $d/\lambda$, $z_0/\lambda$ and ${\bf m}/\phi_0\lambda$ are chosen so that the sequence of panels corresponds temperatures such that  $\lambda/\lambda(0)=1,\,3/2,\,3,6$. Note that in this case the mesh parameter in Figs.\ \ref{fig.fig7} and \ref{fig.fig8}, $a\sim \xi$, also changes with $T$, since $\xi$ is temperature dependent. 

\noindent ${\bf m}$ {\it perpendicular to the film surfaces}: The equilibrium vortex configurations consist of vortex lines with a single flux quantum, that is $\mid n_x\mid=\mid n_z\mid =1$ along the lines, and with $n_z$ of the same sign as $m_z$ (positive in Fig.\ \ref{fig.fig7}). The vortex lines are curved, except for the thinner film with 
$d=0.5\lambda$. The curvature results from the competition between vortex-vortex and vortex-dipole interactions. Closer to the $z=0$ surface the vortex-dipole interaction dominates, pulling the vortex lines towards the dipole, and keeping them perpendicular to the film surfaces. Deeper inside the film  the vortex-dipole interaction weakens and vortex-vortex repulsion separates the vortex lines apart.   The creation of net vorticity by the magnetic dipole obtained here is particular to the isolated dipole. For a dipole array the net vorticity must vanish, even if the dipoles are far apart, due to the long range of vortex-vortex interaction in films \cite{vvlr}.

\noindent ${\bf m}$ {\it parallel  to the film surfaces}: The vortex configurations consist of half loops and pairs of vortex lines with opposite vorticity, both with a single flux quantum, as shown in Fig.\ \ref{fig.fig8}. These configurations reflect the properties of the pinning potential for ${\bf m}$  parallel  to the film surfaces shown in  Figs.\ \ref{fig.fig3}, and \ref{fig.fig4}. For instance, in the case of the half loop for $d=3.0\lambda$, the resulting curve places the negative (positive) $n_z$ in regions of positive (negative) pinning.  In the case of vortex lines, the nearly straight curve follows, essentially the pinning potential maxima for negative $n_z$ and the minima for positive $n_z$. 
The results shown in Fig.\ \ref{fig.fig8} also indicate how the vortices penetrate the film with increasing temperature, that is the transition from the Meissner state to the mixed state. The transition is continuous. Half loops penetrate and grow towards the interior of the film, eventually separating in two lines of apposite vorticities.   

The equilibrium vortex configurations described above, for both orientations of ${\bf m}$, are not expected to change if the restriction that they are in the plane of the dipole is lifted, since there would be no gain in the vortex-vortex or vortex-dipole interaction energies if some vortex lines were out of the $x-z$ plane.  

\section{conclusion}
\label{sec.conc}

In conclusion then this paper solves exactly in the London limit the problem of  vortices in a film of arbitrary thickness interacting with a point magnetic dipole outside the film, and obtains from these solutions the vortex-dipole system energy as a functional of the vortex distribution,  which is minimum for the  equilibrium vortex configurations generated by the dipole. The results reported here can be generalized to any distribution of permanent magnetic moments placed outside the film. The numerical method to obtain equilibrium vortex configurations presented  here is of general validity, and can be applied to three-dimensions and to other distributions of dipoles. 

The London limit used in this paper, is expected to  break down for large values of magnetic dipole, because the inhomogeneous magnetic field created by it destroys superconductivity locally in the film. Roughly speaking, London theory is valid as long as the maximum field of the dipole at the film surface nearer to the it is less than the upper critical field, that is , $m/z^3_0 < \phi_0/(2\pi \xi^2)$, or 
$m/(\phi_0\lambda) < (z_0/\lambda)^3(\lambda/\xi)^2/2\pi$. One indication of this break down is the appearance in the  equilibrium vortex configurations of vortex lines separated by distances  $\leq 2\xi$. In the equilibrium vortex configurations shown in  Figs.\ \ref{fig.fig7}, and \ref{fig.fig8} this occurs for ${\bf m}$ perpendicular to the film surfaces in the film with $d=\lambda$, and  for ${\bf m}$ parallel to the film surfaces in the films with $d=\lambda$ and $d=0.5\lambda$.  In both cases the values of $m$ are found to be in agreement with the condition  stated above. Note, however that  only in the immediate vicinity of the film surface the vortex lines separation is $\leq 2\xi$. Deeper inside the film the vortex lines are separated by distances larger than $2\xi$. This can be interpreted as indicating that the regions where the vortex lines separation is  $\leq 2\xi$ are normal, and that in the regions where the separation is larger than $2\xi$ the vortex configurations obtained in the London limit are  reasonable estimates. In the case of bulk superconductors and films under applied magnetic fields, the interpretation along similar lines of London theory results for the vortex configurations generated by the field  lead to a reasonable first order approximation to the vortex phase diagram. The same is believed to be true here. The London theory results described in this paper, and their generalization to distributions of dipoles, can be applied beyond their strict limits of validity to give a first order approximation to vortex behavior in these systems.

\acknowledgments
Research supported in part by the Brazilian agencies CNPq, CAPES, FAPERJ,  and FUJB, and by ICTP/Trieste. The author thanks Daniel H. Dias and Thiago Lobo for assistance with the  software used to visualize vortex configurations.

\appendix
\section{mathematical details}
\label{sec.math}

Here some details of the derivations in Sec.\ \ref{sec.vmd} are given. 
First it is shown that all cross terms in $E_T$ vanish. 

When  Eqs.\ (\ref{eq.bio}) are substituted in  Eq.\ (\ref{eq.etot}) there are several cross terms containing two distinct fields. Here it is shown that these terms vanish.
The first step is to show that there are no cross terms in 
$E_{\rm in}$ with ${\bf b}^{\rm vi}$ and any homogeneous solution of London equation, denoted as  ${\bf b}^{\rm h}$. This term is
\begin{eqnarray}
&E_c = \displaystyle  \int\!\frac{d^2r_{\perp}}{4\pi}    \int^0_{-d}\; dz 
\nonumber \\ 
  &[\lambda^2{\bf \nabla}\times {\bf b}^{\rm vi }\cdot{\bf \nabla}\times {\bf b}^{\rm h }
  +  {\bf b^{\rm vi}}\cdot {\bf b^{\rm h}}]  \,.
\label{eq.ec1}
\end{eqnarray}
Using the identity
\begin{eqnarray}
&{\bf \nabla}\times {\bf b}^{\rm vi }\cdot{\bf \nabla}\times {\bf b}^{\rm h }  = &\nonumber \\
 &{\bf \nabla}\cdot[{\bf b}^{\rm vi }\times ({\bf \nabla}\times 
{\bf b}^{\rm h })]+
{\bf b}^{\rm vi }\cdot[{\bf \nabla}\times({\bf \nabla}\times {\bf b}^{\rm h })]
   \,,
\label{eq.id1}
\end{eqnarray}
and the fact that ${\bf b}^{\rm h }$ satisfies the homogeneous London equation, $E_c$ can be written as
\begin{equation}   
E_c= \lambda^2\int\! \frac{d^2r_{\perp}}{4\pi}\;
 \hat{\bf z}\cdot [{\bf b}^{\rm vi }\times({\bf \nabla}\times {\bf b}^{\rm h })]\mid^0_{-d} \,.
\label{eq.ec2}
\end{equation}
As shown in Ref.\ \onlinecite{gcehb}, ${\bf b}^{\rm vi }$ at the film surfaces $z=0,\,-d$ points in the $z$-direction,  so that the vector product in the integrand has no $z$-component, and $E_c=0$.
This argument eliminates the cross terms with $ {\bf b}^{\rm vi }$ and 
$ {\bf b}^{\rm stray}$, and  with $ {\bf b}^{\rm vi }$ and
 $ {\bf b}^{\rm in}_{\bf m}$. One cross term is left in $E_{\rm in}$ with the fields $ {\bf b}^{\rm stray}$ and  $ {\bf b}^{\rm in}_{\bf m}$. It is shown next that in the case of  a small current loop this term is canceled  out by the cross term with ${\bf b}^{\rm vac}$  and ${\bf b}^{\rm out}_{\bf m}$ in  $E_{\rm out}$. Denoting these terms by $E_{\rm in\, c}$ and  
$ E_{\rm out\, c}$, respectively, it follows  that
  \begin{eqnarray}  
 && E_{\rm in\, c}= \displaystyle\int\!\frac{d^2r_{\perp}}{4\pi} 
  \int^0_{-d}\;dz \times \nonumber \\ 
 &&  [\lambda^2{\bf \nabla}\times {\bf b}^{\rm stray }\cdot{\bf \nabla}\times {\bf b}^{\rm in}_{\bf m}
  +  {\bf b^{\rm stray}}\cdot {\bf b}^{\rm in}_{\bf m}] \label{eq.inc} \,, \\
 && E_{\rm out\, c} = \displaystyle\int \! \frac{d^2r_{\perp}}{4\pi}\times \nonumber \\
&&\displaystyle \Big[ \int^{\infty}_0\! dz\,   + 
  \int^{-d}_{-\infty}\!\! dz\,  \Big] {\bf b}^{\rm vac}\cdot{\bf b}^{\rm out}_{\bf m}\,.
  \label{eq.outc}
  \end{eqnarray} 
Using arguments similar to those leading to  Eq.\ (\ref{eq.ec2}),  $E_{\rm in\, c}$ can be written as
\begin{equation} 
E_{\rm in\, c}=  \frac{\lambda^2}{4\pi}\int\! d^2r_{\perp}
 \hat{\bf z}\cdot [{\bf b}^{\rm stray }\times({\bf \nabla}\times {\bf b}^{\rm in}_{\bf m})]\mid^0_{-d} \,.
\label{eq.inc2}
\end{equation}
From London equation ${\bf \nabla}\times {\bf b}^{\rm in}_{\bf m}= {\bf a}^{\rm in}_{\bf m}/\lambda^2$
(${\bf b}^{\rm in}_{\bf m}={\bf \nabla}\times {\bf a}^{\rm in}_{\bf m}$), so that 
\begin{equation} 
E_{\rm in\, c}=  \frac{1}{4\pi}\int\! d^2r_{\perp}
 \hat{\bf z}\cdot [ {\bf a}^{\rm in}_{\bf m}\times{\bf b}^{\rm stray }]\mid^0_{-d} \,.
\label{eq.vmdi}
\end{equation}
It is possible to write $E_{\rm out\, c}$ in a similar form, using  ${\bf \nabla}\times {\bf b}^{\rm vac}_{\bf m}=0$. After an integration by parts of  Eq.\ (\ref{eq.outc})the result is   
\begin{equation} 
E_{\rm out\, c}= - \frac{1}{4\pi}\int\! d^2r_{\perp}
 \hat{\bf z}\cdot [ {\bf a}^{\rm out}_{\bf m}\times{\bf b}^{\rm vac }]\mid^0_{-d} \,. 
\label{sec.outc2}
\end{equation}
It follows from the continuity of the fields and vector potentials at the film surfaces that $E_{\rm in\, c}+ E_{\rm out\, c}=0$.

In the case of a permanent magnetic dipole there is only a partial cancelation, and  $E_{\rm in\, c}+ E_{\rm out\, c}=- {\bf m}\cdot {\bf b^{vac}}({\bf r}_0)$, as shown next. The energy $E_{\rm out\, c}$ is written in terms of  scalar potentials using the identities 
\begin{eqnarray}
{\bf b}^{\rm vac}\cdot{\bf b}^{\rm out}_m &= &
{\bf \nabla} \Phi \cdot {\bf \nabla} \Phi_{\bf m}
\nonumber \\
&= &{\bf \nabla}\cdot (\Phi{\bf \nabla} \Phi_{\bf m})-\Phi\nabla^2 \Phi_{\bf m}
\nonumber \\
&= &{\bf \nabla}\cdot (\Phi{\bf \nabla} \Phi_{\bf m})+ -4\pi\Phi\, {\bf m}\cdot{\bf \nabla}\, \delta({\bf r}-{\bf r}_0)\nonumber \,.
\label{eq.id2}
\end{eqnarray}
Substituting in Eq.\ (\ref{eq.outc})results
\begin{eqnarray} 
&E_{\rm out\, c}= - {\bf m}\cdot {\bf b^{vac}}({\bf r}_0)-\nonumber \\ 
&\displaystyle \frac{1}{4\pi}\int\! d^2r_{\perp}\,
 \Phi \; [\hat{\bf z}\cdot{\bf b}^{\rm out}_m]^0_{-d} \,. 
\label{eq.outc3}
\end{eqnarray}
The second term in Eq.\ (\ref{eq.outc3}) cancels out $E_{\rm in\, c}$. This can be shown starting from Eq.\ (\ref{eq.inc2}), and using Eq.(22) of Ref.\ \onlinecite{gcehb} for 
${\bf b}^{\rm stray }$, the homogeneous London equation for ${\bf b}^{\rm in}_{\bf m}$, and continuity of the fields at the film surfaces. 
\begin{figure}
\centerline{\includegraphics[scale=0.20]{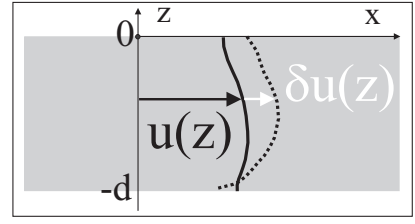}}
\caption{Full line: generic vortex line defined by $u(z)$. Dashed line:  vortex line after deformation $\delta u(z)$. }
\label{fig.fig9}
\end{figure}

\section{work done by the Lorentz force}
\label{sec.wlf}

Here it is shown that the change in the vortex-dipole interaction     energy  $\delta E_{vm}$, Eq.\ (\ref{eq.evm}), equals the negative of the work done on the vortex lines by the Lorentz force of the screening current induced by the magnetic dipole.

Consider the vortex line running from one  film surface to the other shown in Figs.\ \ref{fig.fig9}. The equation describing this line is
\begin{equation}
{\bf r}(z)=z\hat{\bf z}+{\bf u}(z)\;.
\label{eq.vrz}
\end{equation}
The contribution of this line to the vorticity  is 
  \begin{equation} 
  {\mbox{\boldmath $\nu$}}({\bf k})=  \int^0_{-d}dz  \,
  [\hat{\bf z}+\frac{d{\bf u}(z)}{dz}]\,e^{-i({\bf k}_{\perp}\cdot{\bf u}(z) + k_zz)}\, .
  \label{eq.nukl}
  \end{equation}
It is convenient here to work with  Fourier transform  in the $x-y$ plane only. For the vorticity it is 
  \begin{equation} 
  {\mbox{\boldmath $\nu$}}({\bf k}_{\perp},z)=  [\hat{\bf z}+\frac{d{\bf u}(z)}{dz}]\,e^{-i{\bf k}_{\perp}\cdot{\bf u}(z)}\,.  
  \label{eq.nukz}
  \end{equation}
If the vortex line undergoes a small deformation, defined by $\delta{\bf u}(z)$, the change in the vorticity to first order is 
  \begin{eqnarray} 
 &\delta{\mbox{\boldmath $\nu$}}({\bf k}_{\perp},z)= \nonumber \\
 &\displaystyle \{[-i{\bf k}_{\perp}\cdot\delta{\bf u}(z)]\hat{\bf z}+\frac{d\,\delta{\bf u}(z)}{dz}\}\;e^{i{\bf k}_{\perp}\cdot{\bf u}(z) }\,.  
  \label{eq.nukz2}
  \end{eqnarray}
The corresponding change in the vortex-dipole interaction energy is
\begin{eqnarray} 
&\delta E_{vm}/\epsilon_0\lambda= \displaystyle \int\,\frac{d^2k_{\perp}}{(2\pi)^2} \int^0_{-d}\;\frac{dz}{\lambda}\nonumber \\
&\delta\nu_z({\bf k}_{\perp},z)\; U_{vm}(-{\bf k}_{\perp},z)\;.
 \label{eq.devmuz}
  \end{eqnarray}
When the vortex line is deformed by  $\delta{\bf u}(z)$, the Lorentz force of the screening current induced by the magnetic dipole, 
${\bf j}^{sc}=c/4\pi{\mbox{\boldmath ${\nabla}$}}\times{\bf b}^{\rm in}_{\bf m}$, does the work 
\begin{eqnarray} 
\delta W_L&=&\frac{\phi_0}{c}\int\,\frac{d^2k_{\perp}}{(2\pi)^2} \int^0_{-d}\, dz \,{\bf j}^{sc}({\bf k}_{\perp},z)\times\nonumber \\
& & [\hat{\bf z}+\frac{d{\bf u}(z)}{dz}]\cdot\delta{\bf u}(z)\,e^{-i{\bf k}_{\perp}\cdot{\bf u}(z)}\;.
 \label{eq.wlf}
  \end{eqnarray}
The screening current is perpendicular to the z-direction, and is given by (Ref.\ \onlinecite{myp})
\begin{eqnarray} 
&{\bf j}^{sc}({\bf k}_{\perp},z)=\displaystyle\frac{c}{\lambda^2}(i{\bf k}_{\perp}\times \hat{\bf z})(i{\bf m}_{\perp}\cdot{\bf k}_{\perp}+m_zk_{\perp})e^{-k_{\perp}z_0}\times \nonumber \\
&\displaystyle  \frac{(k_{\perp}+\tau)e^{\tau(z+d)} - (k_{\perp}-\tau)e^{-\tau(z+d)}}
{{\cal C}k_{\perp}}\,,
 \label{eq.jsc}
  \end{eqnarray}
Because both the screening current and $\delta{\bf u}(z)$ are parallel to the film surfaces, the term with $d{\bf u}(z)/dz$ in  Eqs.\ (\ref{eq.wlf}) vanishes.
Substituting  Eq.\ (\ref{eq.jsc})in  Eq.\ (\ref{eq.wlf}), and using the expression for $U_{vm}$ obtained in Sec.\ \ref{sec.pinp}, it follows  that 
$\delta E_{vm}=-\delta W_L$. This result can also be demonstrated for vortex lines that cannot be described by  Eqs.\ (\ref{eq.vrz}), such as loops and lines with humps.

\end{document}